\documentclass[nofootinbib,aps,superscriptaddress,preprint,floatfix]{revtex4}

\usepackage{graphicx,epsf}

\usepackage{bm}

\def\to{\rightarrow}
\newcommand{\nc}{\newcommand}
\nc{\beq}{\begin{equation}}
\nc{\eeq}{\end{equation}}
\nc{\barray}{\begin{eqnarray}}
\nc{\earray}{\end{eqnarray}}
\nc{\barrayn}{\begin{eqnarray*}}
\nc{\earrayn}{\end{eqnarray*}}
\nc{\bcenter}{\begin{center}}
\nc{\ecenter}{\end{center}}
\nc{\ket}[1]{| #1 \rangle}
\nc{\bra}[1]{\langle #1 |}
\nc{\mc}{\mathcal}
\nc{\er}[1]{(\ref{eq:#1})}
\nc{\onehalf}{\frac{1}{2}}
\nc{\partialbar}{\bar{\partial}}
\nc{\psit}{\widetilde{\psi}}
\nc{\Tr}{\mbox{Tr}}
\nc{\hc}{\mbox{H.c.}}
\nc{\ev}{\;\mathrm{eV}}
\nc{\mev}{\;\mathrm{MeV}}
\nc{\gev}{\;\mathrm{GeV}}
\nc{\tev}{\;\mathrm{TeV}}

\def\chii0{\chi_i^0}
\def\chij0{\chi_j^0}

\newcommand{\gsim}{\lower.7ex\hbox{$\;\stackrel{\textstyle>}{\sim}\;$}}
\newcommand{\lsim}{\lower.7ex\hbox{$\;\stackrel{\textstyle<}{\sim}\;$}}
\nc{\ttbar}{t\bar t}

\begin{document}


\preprint{SLAC-PUB-14408}
\preprint{UCI-HEP-TR-2011-05}

\title{\bf \Large ${\mathbf A_{FB}^t}$ Meets LHC
}

\author{\large JoAnne L. Hewett}
\affiliation{SLAC National Accelerator Laboratory, 2575 Sand Hill Road, Menlo Park, CA 94025, USA}

\author{\large Jessie Shelton}
\affiliation{Department of Physics, Yale University, New Haven, CT 06511, USA}

\author{\large Michael Spannowsky}
\affiliation{Institute of Theoretical Science, University of Oregon, Eugene, OR 97403-5203, USA}

\author{\large Tim M.P. Tait}
\affiliation{Department of Physics and Astronomy, University of California, Irvine, CA 92697, USA}

\author{\large Michihisa Takeuchi}
\affiliation{Institut f\"{u}r Theoretische Physik, Universit\"{a}t Heidelberg, Germany}

\vskip 0.5in

\begin{abstract}

The recent Tevatron measurement of the forward-backward asymmetry of the top quark shows
an intriguing discrepancy with Standard Model expectations, particularly at large $\ttbar$
invariant masses. Measurements of this quantity are subtle at the LHC, due to its
$pp$ initial state, however, one can define a forward-central-charge asymmetry which captures the physics.
We study the capability of the LHC to measure this asymmetry and find that within the SM
a measurement at the $5\sigma$ level is possible with roughly 60 fb$^{-1}$
at $\sqrt{s} = 14$~TeV.  If nature
realizes a model which enhances the asymmetry (as is necessary to explain the Tevatron
measurements), a significant difference from zero can be observed much earlier, perhaps
even during early LHC running at $\sqrt{s} = 7$~TeV.  We further explore the capabilities
of the 7 TeV LHC to
discover resonances or contact interactions which modify the $\ttbar$ invariant
mass distribution using recent boosted top tagging techniques.  We find that TeV-scale color octet
resonances can be discovered, even with small coupling strengths and that contact
interactions can be probed at scales exceeding 6 TeV.
Overall, the LHC has good potential to clarify the situation with regards to the Tevatron forward-backward
measurement.

\end{abstract}

\def\thepage{{}}
\maketitle
\def\thepage{\arabic{page}}

\section{Introduction}

The top quark, with its large mass, represents a perfect laboratory to probe
electroweak symmetry breaking and its associated new physics.
The Tevatron program has produced thousands of
top quarks and overall observes good agreement between measured top properties and
 the expectations of the Standard Model (SM).  As the LHC accumulates luminosity, there is
 potential to study the top in a whole new energy regime, perhaps revealing secrets that
 are too subtle to discern at lower energies.

One particular measurement at the Tevatron already shows an intriguing deviation from
the SM expectation, the top-quark forward-backward asymmetry
$A ^ t_{FB}$.  This observable measures the tendency of the top quark in a $\ttbar$ pair
to move forward along the same direction as the incoming quark as opposed to backward
in the direction of the incoming anti-quark.  In the Standard Model, this quantity deviates from
zero at next-to-leading order (NLO) in QCD, and thus can be a sensitive probe of new physics which
contributes in principle at tree level.  Previous measurements by both
CDF and D0 showed a tantalizing (but not very significant) discrepancy with the SM
expectation \cite{:2007qb,Aaltonen:2008hc,Aaltonen:2009iz}.  The situation has
become even more fascinating with the recent CDF study based on larger statistics and
examining separately the high and low $\ttbar$ invariant mass ($M_{\ttbar}$)
regions \cite{Aaltonen:2011kc}.
At low invariant masses, the asymmetry is roughly consistent with the Standard Model
expectations.  At high invariant masses $M_{\ttbar} >450\gev$, where one might
naturally expect the effects of heavy new physics to manifest, the asymmetry is
observed to be nearly $50\%$, exceeding the SM prediction by $3.4\sigma$.

This result is striking, and emphasizes the need
to better understand the top forward-backward asymmetry within the SM, to explore
models of physics beyond the Standard Model (BSM) that could potentially explain the
Tevatron measurements, and for further experimental data.  In particular, the LHC is a natural
place to explore high energy top physics, and could potentially provide very instructive data.
However, the measurement of $A_{FB}^t$ is somewhat subtle at the LHC, because its
charge-symmetric $pp$ initial state does not provide an automatic bias on the initial
state partons participating in the reaction.  Indeed, for the typical $gg$ initial state mainly responsible
for $\ttbar$ production at the LHC, there is no asymmetry at all.  One is thus forced to rely
on the subdominant $q\bar q$ and $qg$ initial states, making use of the strong correlation
between the rapidity of the $\ttbar$ system and the direction of the incoming
quark.  Event-by-event, a positive
forward-backward asymmetry in $q\bar q (qg) \to \ttbar + X$ translates into a positive
charge asymmetry in the forward regions $|y| > y_0 $ \cite{KuhnandRodrigo, Antunano:2007da}.
This {\it forward-central} charge asymmetry is (assuming CP conservation)
a natural LHC observable.\footnote{There have been several other recent
proposals for interesting related observables \cite{Antunano:2007da,WangXiaoZhu,Xiao:2011kp,Choudhury:2010cd,Rodrigo:2010gm}.}

We will study the LHC prospects for measuring this top forward-central charge asymmetry, for both
SM and BSM theories designed to explain the Tevatron measurement.
We will see that 60 fb$^{-1} $ at 14 TeV will allow for the small SM asymmetry to
be observed, while larger asymmetries from models of new physics could be measured much sooner with tens of fb$^{-1} $ at 7 TeV.

Ultimately, models which aim to explain the Tevatron measurements
\cite{Jung:2009jz,Frampton:2009rk,Shu:2009xf,Djouadi:2009nb,Arhrib:2009hu,Cheung:2009ch,Barger:2010mw,
Cao:2010zb,Alvarez:2010js,Shelton:2011hq,Barger:2011ih,Grinstein:2011yv,Patel:2011eh,Isidori:2011dp}
may manifest themselves in a variety of ways at the LHC.
In addition to modifications of $A_{FB}^t$,
virtually all of the candidate theories result in
an increase in $\ttbar$ production at large invariant mass, $M_{\ttbar}$, and the
differential cross section $d\sigma_{\ttbar}/dM_{\ttbar}$ can be a sensitive discriminant between
various models \cite{Bai:2011ed,Bhattacherjee:2011nr,Xiao:2010hm}.
In fact, the unprecedented kinematic reach afforded by the LHC may
allow production of new states which could contribute only virtually to Tevatron processes,
revealing new $\ttbar$ resonances.  To this end, we
investigate the LHC sensitivity to both resonant and non-resonant modification of
the $\ttbar$ invariant mass spectrum.\footnote{
Some models may also
produce even more unusual signals of top physics such as
same sign tops \cite{SSTops} and resonances decaying into a top quark and a light
unflavored jet  \cite{Gresham:2011dg}.}

The organization of this paper is as follows.  In section~\ref{sec:afb}, we introduce a
statistically stable definition of the integrated forward-central charge asymmetry
and study the LHC sensitivity at $\sqrt s= 7$ and 14 TeV.  In sections~\ref{sec:resonance} and
\ref{sec:contact}, we consider search prospects for resonances and contact operations in
early LHC data.  Section~\ref{sec:conclusion} contains our conclusions.

\section{Measuring $A ^ t_{FB}$ at the LHC}
\label{sec:afb}

As mentioned above, the forward-central charge asymmetry is a natural LHC observable
probing the forward-backward asymmetry in $\ttbar $ pair production.
It proceeds by dividing top (and anti-top) quarks based on their
rapidities $y$ between central and forward regions of the
detector.  The division in rapidity, $y_0$
defines the forward $|y|>y_0$ and central $|y|<y_0$ regions, and can be chosen
to provide optimum sensitivity to the asymmetry.

For a given $y_0$, we define both a forward charge asymmetry,
\beq
\label{eq:asymmetry1}
\mathcal{A}_F (y_0) =\frac{N_t (y_0 < | y | < 2.5)-N_{\bar t}
(y_0 < | y | < 2.5)}{N_t (y_0 < | y | < 2.5)+N_{\bar t }(y_0 < | y| < 2.5)},
\eeq
and a central charge asymmetry,
\begin{eqnarray}
\label{eq:asymmetry2}
{\cal A}_C(y_0)=\frac{N_{t}(|y|<y_0) - N_{\bar{t}}(|y|<y_0) }{N_{t}(|y|<y_0) + N_{\bar{t}}(|y|<y_0)}.
\end{eqnarray}
Both definitions exploit the fact that the quark parton distribution functions
(PDF) have more support at large parton $x$ than
either the gluon or anti-quark PDFs,
resulting in an event-by-event
correlation between the rapidity of the $\ttbar$ pair and the incoming quark direction.
As a result,
a positive forward-backward asymmetry implies that the number of anti-top
quarks in the central region is larger than the the number of top quarks, while the
total number of top and anti-top quarks integrated over the whole rapidity region is the same
(up to finite $\eta$ acceptance, which we find to be a negligibly small effect).
Thus, ${\cal A}_F$ and ${\cal A}_C$ will have opposite signs.
Note that in a given event both the top and the anti-top can be either central or forward;
with the definitions of Eq.(\ref{eq:asymmetry1}) and Eq.(\ref{eq:asymmetry2}), a single event
can thus contribute to both ${\cal A}_F$ and ${\cal A}_C$, and the two observables are not independent.
Since the central region contains a larger proportion of symmetric $gg$ initiated $\ttbar$ events,
the forward charge asymmetry $\mathcal{A}_F (y_0)$ is a more sensitive probe of the underlying asymmetry
in the $\ttbar$ cross-section.

To estimate the potential to measure a charge asymmetry with a specific significance we define
the significance of an asymmetry observable as,
\begin{eqnarray}
\label{eq:sig}
\sigma_{\cal A}(y)=\frac{|{\cal A}(y)|}{\Delta {\cal A}(y)},
\end{eqnarray}
where $\Delta {\cal A}(y)$ is the statistical uncertainty on ${\cal A}(y)$
\begin{eqnarray}
\Delta {\cal A}(y)=\frac{\sqrt{[\Delta N_{t}]^2 + [\Delta N_{\bar{t}}]^2} }{N_{t} + N_{\bar{t}}}.
\end{eqnarray}
In this study we confine ourselves to estimates including statistical uncertainties.
Systematics may prove important as well, but require detailed detector simulations which
are beyond the scope of this work.

\subsection{Simulations}

To generate the Standard Model signal we use MC@NLO \cite{Frixione:2002ik}
and shower those events with Herwig. We normalize the $t \bar{t}$ production cross section for
$\sqrt{s}=14~\rm{TeV}$ to its SM NNLO rate,
$\sigma_{14~\rm{TeV},NNLO}=918~\rm{pb}$ \cite{Moch:2008qy}
and for $\sqrt{s}=7~\rm{TeV}$ to the NLO cross-section obtained from
MC@NLO, $\sigma_{7~\rm{TeV},NLO}=150~\rm{pb}$. After the selection cuts described below,
the only major background is $W+$ jets, which we generate using
Alpgen \cite{Mangano:2002ea}, applying the
MLM matching procedure for up to 5 jets \cite{Hoche:2006ph}.  CTEQ6M PDFs
\cite{Nadolsky:2008zw} are used for the NLO SM processes.

As a representative BSM model which can explain the Tevatron measurement of the
forward-backward asymmetry, we choose the flavor-violating $Z^\prime$ model of
 \cite{Jung:2009jz} at its ``best fit" parameters of
 $\alpha_X = 0.024$ and $M_{Z^\prime} = 160$~GeV.  While this is just one model
 capable of explaining the Tevatron measurement, and its LHC predictions are not
 particularly representative of the range of possibilities, it does provide a well-motivated
 example for how the asymmetry could turn out at the LHC if the current Tevatron measurements
 are in fact the result of physics beyond the Standard Model.  We generate $\ttbar$ production
 in this model at leading order using MadEvent \cite{Alwall:2007st}
 (with CTEQ6L PDFs) showered using Pythia \cite{Sjostrand:2006za}.  It is worth mentioning
 that should this model actually turn out to be realized, there will be like-sign top and
 resonances decaying into a top and a light quark
 which may be observable with smaller data sets than will
 be necessary to measure a significant asymmetry at the LHC 
\cite{Jung:2009jz,SSTops,Gresham:2011dg}.

\subsection{Analysis} 
\label{sec:analysis}

Our analysis looks for $\ttbar$ where one top decays semi-leptonically (allowing a
determination of the parent top charge) and the other
hadronically.  These ``lepton + jets" events strike a balance between the desire to identify
the charge of the top through its leptonic decay
with a larger branching ratio and thus better statistics
on the measurement from the hadronically decaying top.
We proceed by selecting
events with exactly one isolated lepton with $p_T > 15~\rm{GeV}$ and $|y|<2.5$.
A lepton is considered isolated provided the hadronic transverse energy
$E_T$ in a cone of $R=0.3$ around the lepton is less than $30\%$ of the lepton's
transverse energy $E_{T,l}$, i.e. $E_{T}/E_{T,l} < 0.3$.

A non-zero asymmetry is entirely the result of $q\bar q$ and $q g$ initial states, and it is desirable
to impose cuts which select these compared to the dominant $g g$ initial state (which
exhibits no asymmetry, and thus washes out the measurement).  The
fact that the valence quark PDFs have larger support at large parton $x$ than the gluon PDF
proves to be useful.  Requiring large $t \bar{t}$ invariant mass results in a sample with
a relatively  greater proportion of $q\bar q$ and $q g$ initial states, and thus a larger
charge asymmetry.
We enforce
large $M_{\ttbar}$ by requiring the tops to have large transverse momentum.  This has the
drawback of reducing the acceptance, and smaller statistics available for
the measurement.
However, this price is offset by the fact that
tops at moderate boost ($ p_T \gsim m_t $) allow for cleaner
event reconstruction.  In particular, combinatoric backgrounds are reduced,
improving the reconstruction of the parent top rapidity.   Modestly boosted tops
also offer improvements in $b$-tagging efficiency and acceptance of daughter $W$ bosons.
Such modestly boosted (hadronically decaying) top quarks
are well suited for reconstruction using the
HEPTopTagger, which is designed
to operate on tops with $ p_T \gsim 200\gev $ \cite{Plehn:2009rk,Plehn:2010st}.

After removing the isolated lepton, we group all remaining visible particles into massless cells of
size $\Delta \eta \times \Delta \phi = 0.1 \times 0.1$.
Cells with transverse energy above 0.5 GeV are retained for jet clustering.
Initially, the event is clustered (using FastJet \cite{Cacciari:2005hq}) according to
the Cambridge-Aachen jet-finding algorithm \cite{Dokshitzer:1997in}  with a large
effective cone size $R=1.5$.  We require that the hardest jet have transverse momentum larger
than $p_{T,\rm{min}}>200~\rm{GeV}$; we will refer to this jet as the ``fat jet".  Harder $p_T$ cuts
result in too much loss of acceptance and ultimately
reduce sensitivity to the asymmetry. We require this jet to be tagged as a top candidate by the
HEPTopTagger \cite{Plehn:2009rk,Plehn:2010st}. To suppress
the $W+$jets background we additionally require one $b$-tag
in the fat jet, for which we assume a $60\%$ tagging efficiency and a
$2\%$ fake rate. Note that at the LHC the $W+$jets background produces
more positively charged leptons than negatively charged leptons. This can fake a bigger
positive charge asymmetry than produced from the SM $\ttbar$ process alone, and
it is critical to reduce this background in order to accurately determine the top charge asymmetry.
After requiring the hadronic top to be
reconstructed using the HEPTopTagger and demanding the $b$ tag,
we find that the $W+$ jets background becomes irrelevant, see Table \ref{tab:14tev}.

The asymmetries of Eq.(\ref{eq:asymmetry1}) and Eq.(\ref{eq:asymmetry2}) depend on the
rapidity of the leptonic top as well.
Several approaches have been proposed for this purpose \cite{Rehermann:2010vq, Plehn:2011tf}.
Here we take the simple approach of identifying the
leptonic top rapidity with the lepton rapidity.
The degree of the correlation of the lepton rapidity with that of its parent top depends
to some degree on the top polarization \cite{Tait:2000sh} and in principle one could recover additional
information from a more careful treatment of the semileptonic top.
While the degree of polarization increases with increasing boost, so
does the kinematic correlation of the lepton rapidity with the
top rapidity.  In our present study, as we have not retained
polarization information in modeling the top decays, the lepton
rapidity provides a good approximation to the top rapidity regardless
of the production mechanism.

A further cut can improve the statistical significance of the asymmetry. After identifying the region with optimal $\mathcal{A}_F$ and $\mathcal{A}_C$ we veto events where both tops are either forward or central. Those events do not contribute to the asymmetry. Although this cut results only in a small increase of the statistical significance to extract the forward asymmetry $\mathcal{A}_F$, it is very effective in removing symmetric gluon-gluon initial states. Thus, by definition, $\mathcal{A}_F$ and $\mathcal{A}_C$ contain the same information and become equally significant.

\subsection{Results}

\begin{figure}[h]
\includegraphics[width=6.0cm]{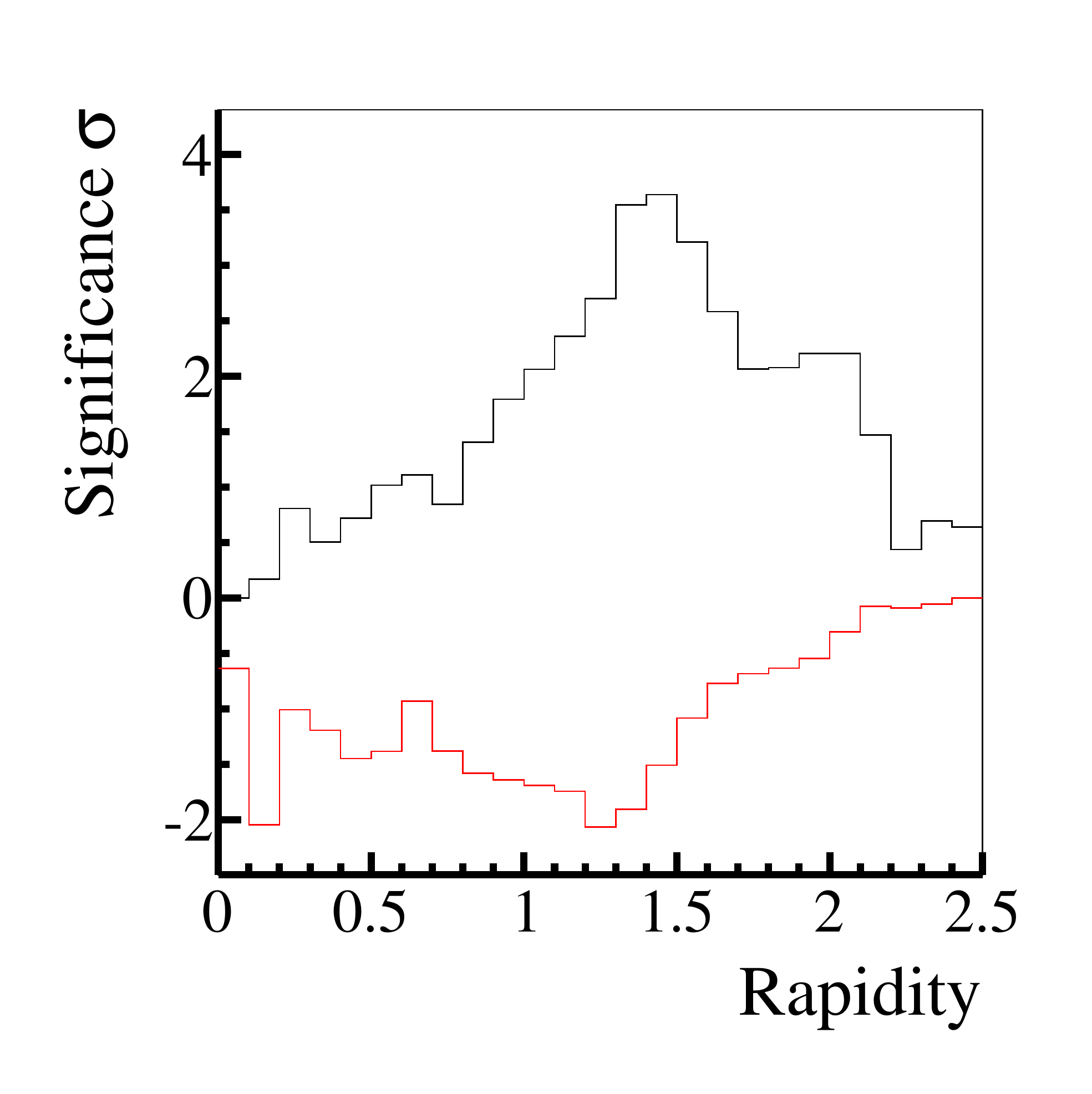}
\includegraphics[width=6.0cm]{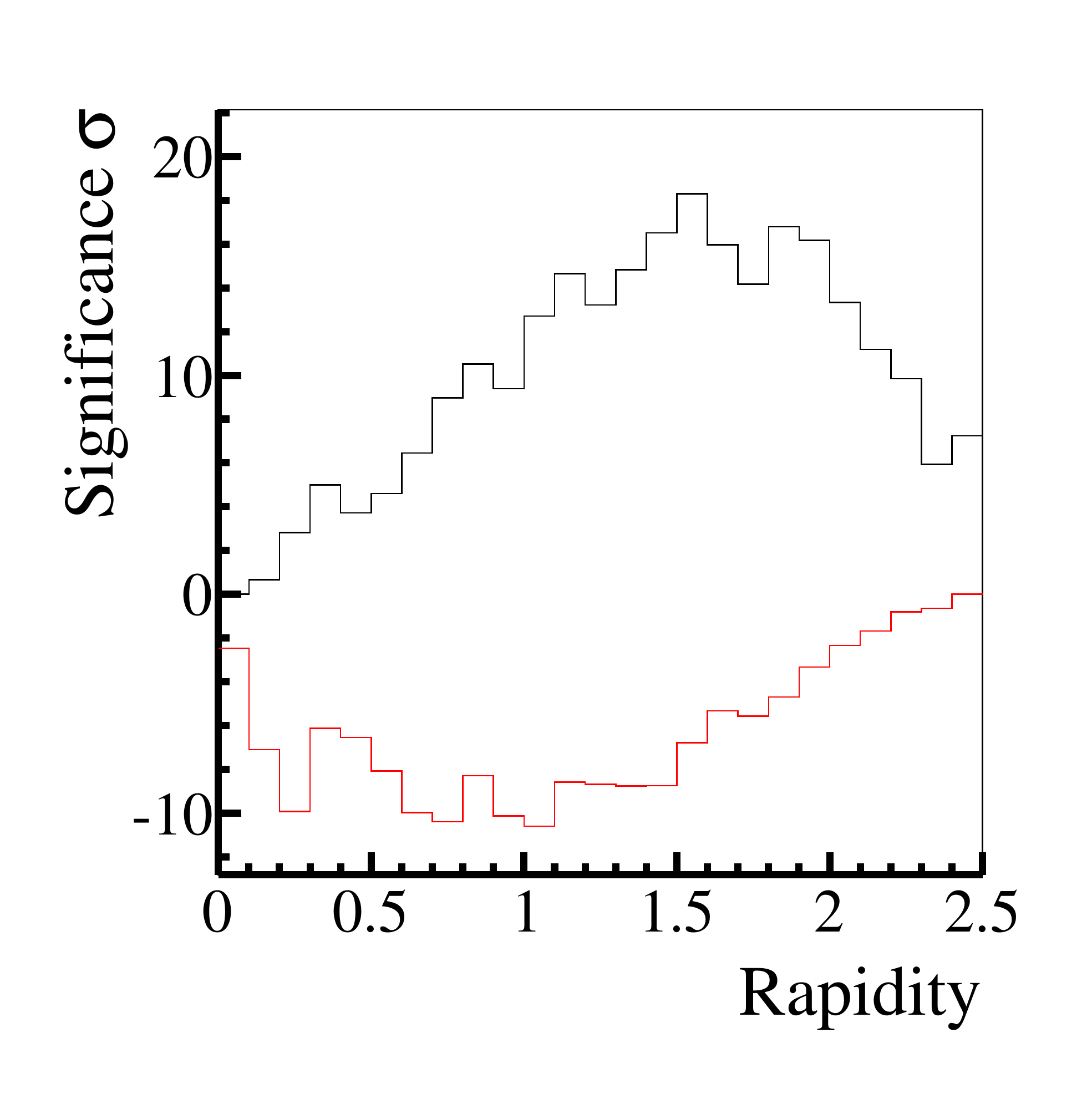} \\
\includegraphics[width=6.0cm]{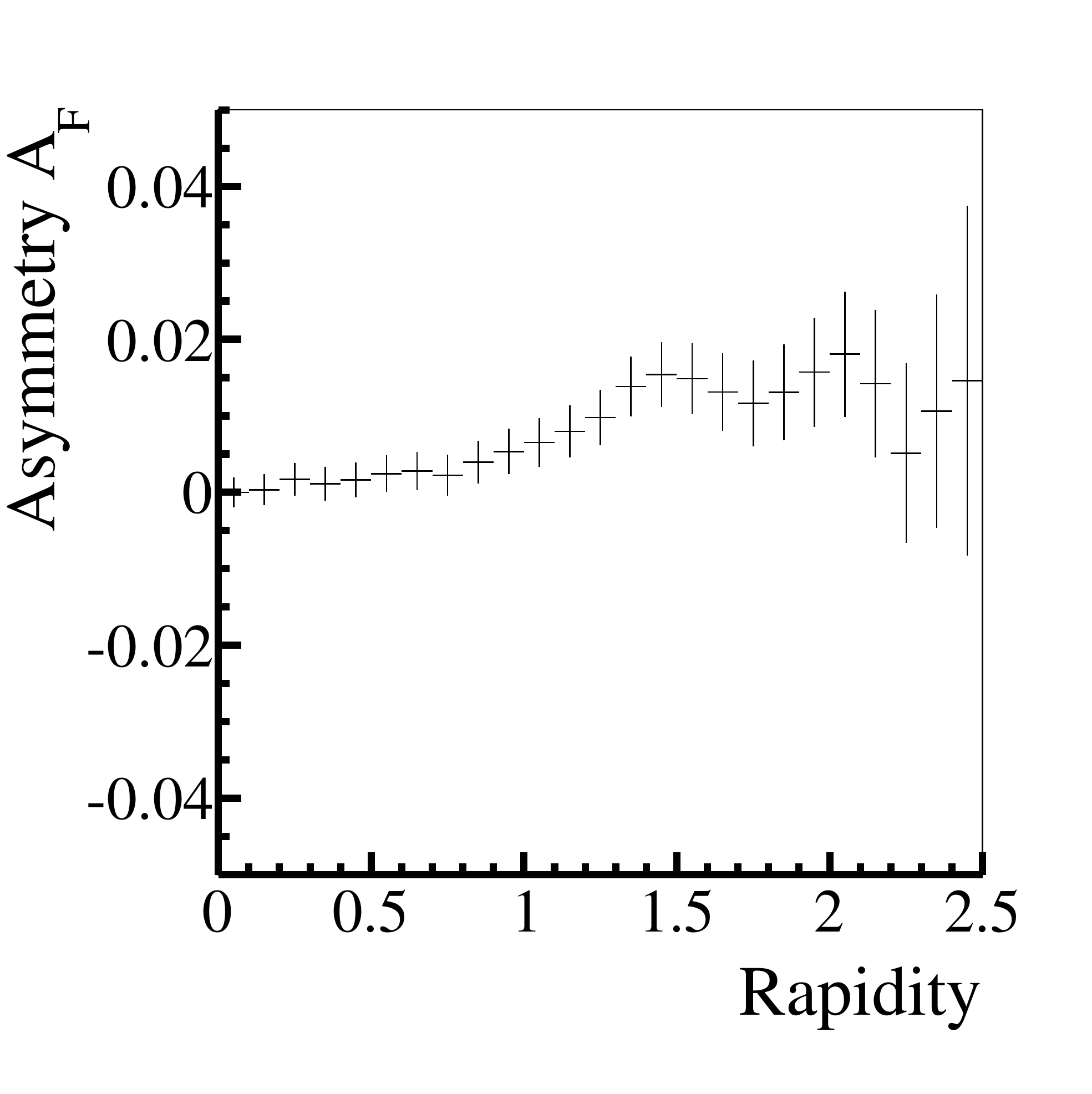}
\includegraphics[width=6.0cm]{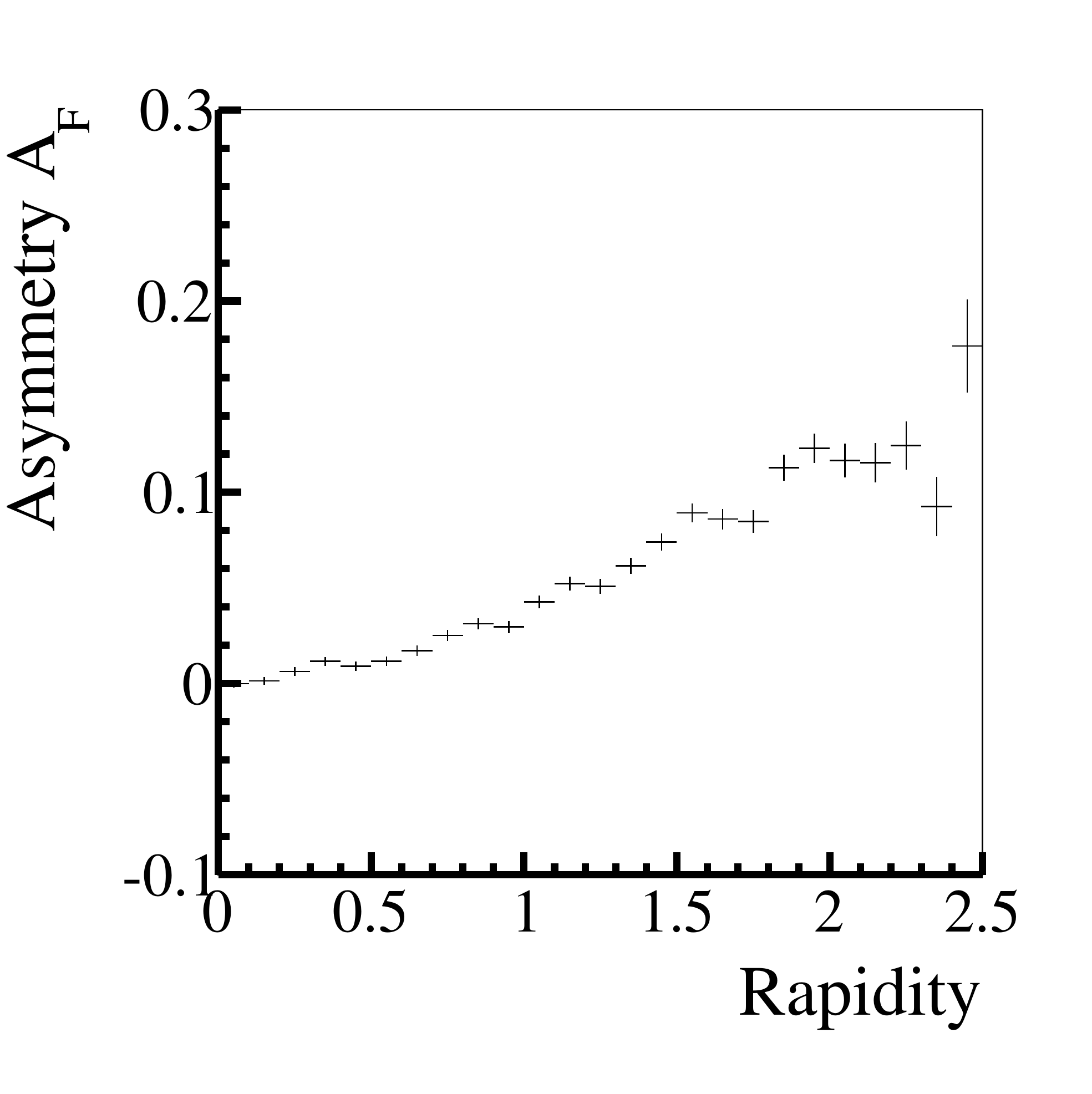}
\includegraphics[width=6.0cm]{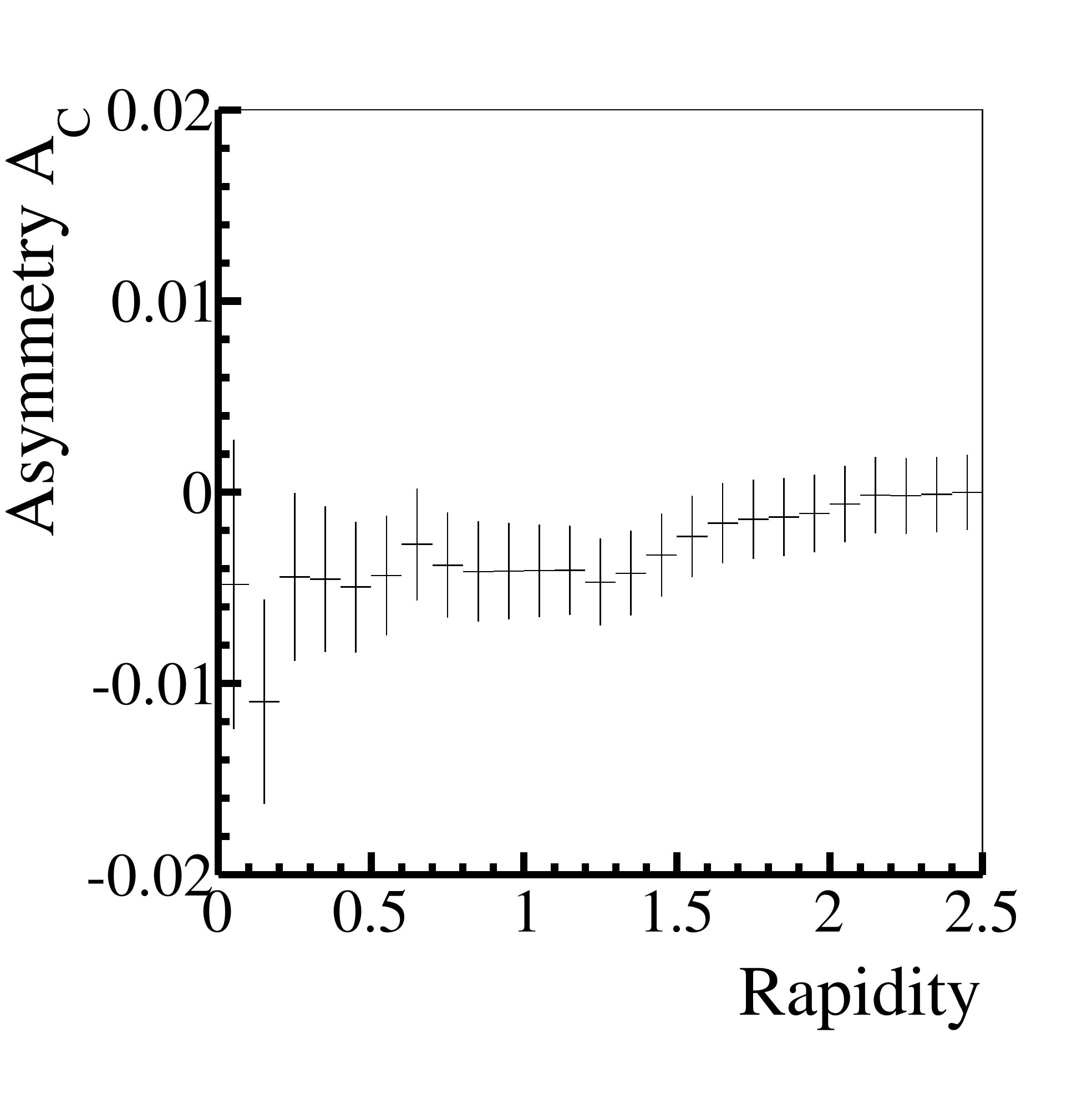}
\includegraphics[width=6.0cm]{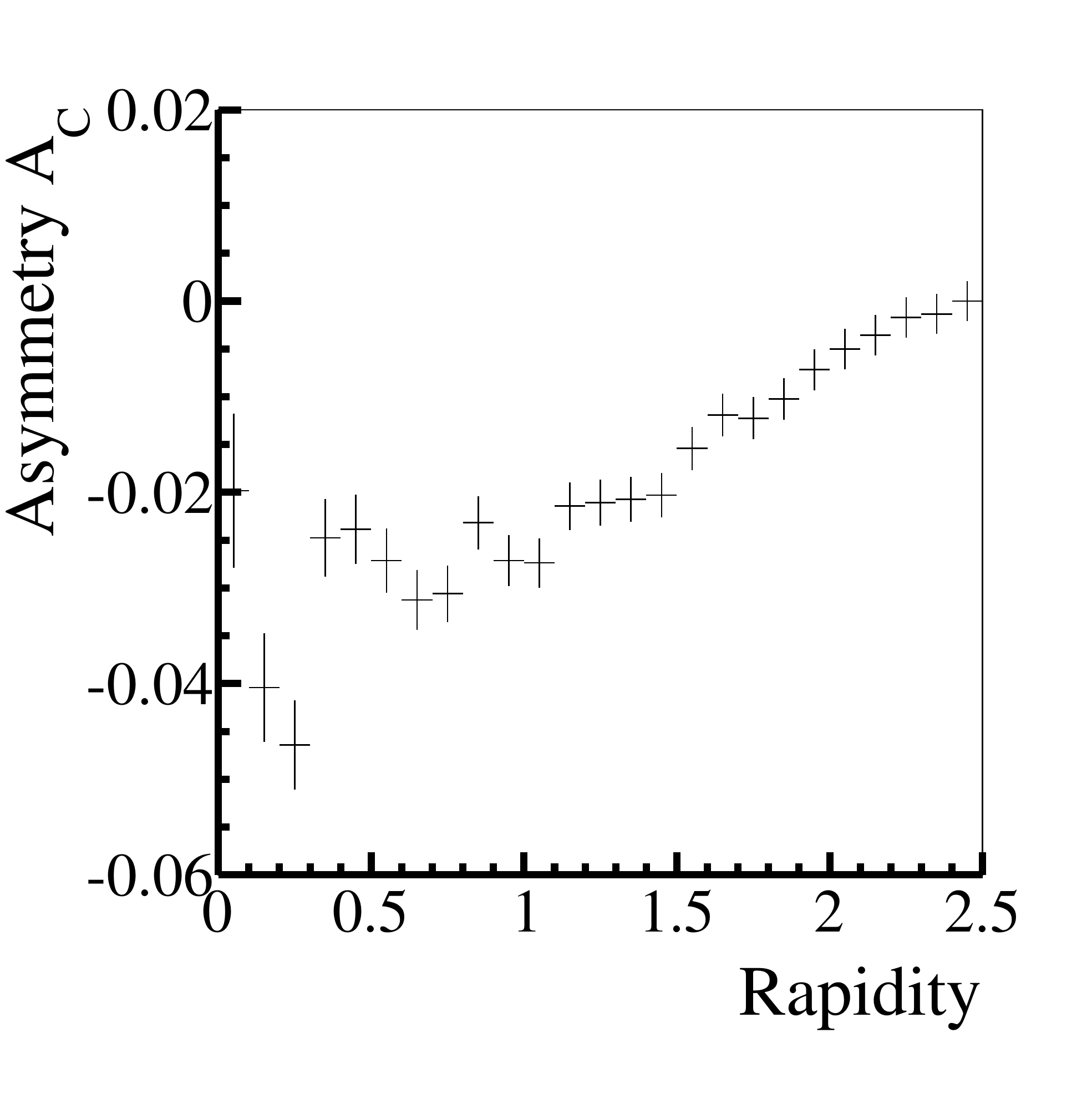}
\caption{The upper panels show the asymmetry significance of Eq.(\ref{eq:sig}) as a function
of $y_0$, for $\mathcal{A}_F (y_0)$ (black) and $\mathcal{A}_C (y_0)$ (red).  The left plot shows
results for the SM and the right plot for the $Z^\prime$ model
described in the text. The middle panels show $\mathcal{A}_F (y_0)$ for SM (left) and BSM (right)
as a function of $y_0$ including statistical error bars. The lower panels show
$\mathcal{A}_C (y_0)$ for SM (left) and BSM (right) as a function of $y_0$ including statistical
error bars.
\label{fig:y0}}
\end{figure}

Our results are summarized in Figure~\ref{fig:y0} and in Tables~\ref{tab:14tev}
and \ref{tab:7tev}.  Figure~\ref{fig:y0} shows the asymmetry and its
significance as a function of $y_0$ for both the SM and $Z^\prime$ models.  As expected,
${\cal A}_F$ is more sensitive than ${\cal A}_C$.  We find that an optimum choice
for the separation parameter is $y_0 \sim 1.5$, which we adopt in deriving the numbers in
the tables.
Tables~\ref{tab:14tev} ($\sqrt{s} = 14$~TeV)
and \ref{tab:7tev} ($\sqrt{s} = 7$~TeV), show
the number of events per fb$^{-1}$, resulting ${\cal A}_{F,C}$ for 25 (10) fb$^{-1}$,
and the statistical significance as defined in Eq.~(\ref{eq:sig}) of the deviation of ${\cal A}$ from
zero for both the SM and our reference $Z^\prime$ model.

We conclude that the SM forward charge asymmetry can reach the
$5\sigma$ level at the LHC with $\sqrt{s}=14$ TeV after about 60 ${\rm fb}^{-1}$.
If the charge asymmetry is larger, compatible with the effect measured by CDF, the asymmetry can
usually be extracted much earlier. For the $Z^\prime$ model introduced in \cite{Jung:2009jz} a
$5\sigma$ discovery can be achieved after less than 2 ${\rm fb}^{-1}$.
With early data at $\sqrt{s}=7$ TeV it will be challenging to extract the charge asymmetry
reliably. $2-3 \sigma$ evidence might  be observable after $10~{\rm fb}^{-1}$,
yielding $2\sigma$ at 5 ${\rm fb}^{-1}$ if it is enhanced by new physics, but the SM asymmetry
cannot be reconstructed from early data using boosted top quarks.
If we require for the accepted events that one of the tops has to be in the central region and one in the forward region,  where $|y|=1.5$ discriminates between those two regions, we find a small increase of the statistical significance for all of the scenarios. After $25~\rm{fb}^{-1}$ at $\sqrt{s} = 14$~TeV we find for the SM asymmetry $3.65\sigma$ and for the $Z^\prime$ model $20.9\sigma$. At $\sqrt{s} = 7~\rm{TeV}$ with $10~\rm{fb}^{-1}$ we find for the SM $0.38\sigma$ and for the $Z^\prime$ model $3.0\sigma$.
It would be interesting to try top reconstruction methods which are effective at
lower $p_T$ \cite{Kondo:1988yd, Barger:2011wf}, but such studies are beyond the scope of
this work.

\begin{table}[t]
\begin{tabular}{ll|ccc||r|rr}
 &   & $(N_+ + N_-)$\,, &$(N_+ - N_-)$\,, & $\Delta N$ & ${\cal
A}_{(F,C) 25\text{fb}^{
-1}}\ \ \ \ \ $ & $\sigma_{25\text{fb}^{-1}}$\cr
\hline
$t\bar{t}$ ~(SM 14~TeV) ~
&$ 0<|y|<2.5$ & 10350  & 0 & &  & &\cr
&$1.5<|y|<2.5$ & 1869  & 27.8 & 43.2&  $0.0149 \pm 0.0046$ & 3.24 \cr
&$ 0<|y|<1.5$ &  8481 & -27.8 & 92.1&$-0.0033 \pm 0.0022$& 1.51 \cr
\hline
\hline
$t\bar{t}$ ~(BSM 14~TeV) ~
&$ 0<|y|<2.5$ &9153 & 0 &  && \cr
&$1.5<|y|<2.5$ &1695 & 151.3 & 41.2 & $0.0893 \pm 0.0049$ & 18.3 \cr
&$ 0<|y|<1.5$ & 7458 & -151.3 & 86.4&$-0.0203 \pm 0.0023$&  8.76 \cr
\hline
\hline
$W$+jets ~(SM 14~TeV) ~
&$ 0<|y|<2.5$ & 20.88 & 0 & & & \cr
&$1.5<|y|<2.5$ &5.54 & 0.46 & 2.35 &    - & - \cr
&$ 0<|y|<1.5$ &15.34 & -0.46 & 3.92 & - & - \cr
\hline
\end{tabular}
\caption{Expected number of events after an integrated luminosity of 1 fb$^{-1}$
at $\sqrt{s} = 14$~TeV. The total cross section is normalized to 918pb (NNLO). We give the error
for the resulting asymmetry and its significance, $\sigma$, for 25 fb$^{-1}$.}
\label{tab:14tev}
\end{table}
\begin{table}[th!]\begin{tabular}{ll|ccc||r|rr}
 &   & $(N_+ + N_-)$\,, &$(N_+ - N_-)$\,, & $\Delta N$ & ${\cal A}_
{(F,C)10\text{fb}^{-1}}
\ \ \ \ \ $ & $\sigma_{10\text{fb}^{-1}}$\cr
\hline
$t\bar{t}$ ~(SM 7~TeV)~
&$ 0<|y|<2.5$ & 1390.9 & 0 & & & \cr
&$1.5<|y|<2.5$ &163.8 & 1.43 & 12.8 & $0.0087 \pm 0.025$ & 0.35 \cr
&$ 0<|y|<1.5$ & 1227.1  & -1.43 & 35.0& $-0.0012 \pm 0.0090$& 0.13\cr
\hline
\hline
$t\bar{t}$ ~(BSM 7~TeV) ~
&$ 0<|y|<2.5$ & 1194.8 & 0 & & & \cr
&$1.5<|y|<2.5$ &140.3 & 10.55 & 11.8 &     $0.075 \pm 0.027$ & 2.81 \cr
&$ 0<|y|<1.5$ &1054.5 & -10.55 & 32.5 & $ -0.010 \pm 0.0097 $& 1.03\cr
\hline
\end{tabular}
\caption{Expected number of events after an integrated luminosity of 1 fb$^{-1}$. The total cross section is normalized to $150$ pb (NLO). We give the error
 for the resulting asymmetry and its significance, $\sigma$, for 10 fb$^{-1}$.}
\label{tab:7tev}
\end{table}

\section{Resonances}
\label{sec:resonance}

While the top-quark forward-central asymmetry will take significant integrated luminosity to detect,
new physics in the top invariant mass spectrum can be observed in the early phases of LHC running with center-of-mass
energies of 7 TeV.
The intermediate mass range, $1 \tev \lsim M_{\ttbar} \lsim 2 \tev$, is of special interest,
both because the LHC will be able to quickly probe this region, and because the Tevatron anomaly suggests
it is a prime spot for new physics to appear.  This intermediate mass range is also a particularly
interesting kinematic environment.  Top-quark pairs with invariant mass in this range
will typically have some, but not all, of their decay products merging
in the detector.  To fully utilize the events in this range, it is necessary to use techniques that interpolate between
traditional top reconstruction methods using isolated jets and leptons \cite{oldATLAS},
and new approaches designed to tag the top as a single object (see \cite{Boost2010} for a recent review).
The ultimate performance of these new reconstruction techniques is still to be
validated, and here we focus on obtaining representative estimates of sensitivity to the production
of new resonances in the $t\bar t$ channel,
rather than on furthering the tagging state-of-the-art.

We compute the resonance search reach for various generic models using
the semileptonic channel resulting from top-quark pair production.
We adopt the top tagger described in \cite{ATLAStopres}, which is based on the work
in Refs. \cite{Kaplan:2008ie,moretops}.  The baseline version of this tagging technique yields the
best efficiency at lower top-quark momentum and we thus employ that version here.
This reconstruction technique has been specifically designed for
top-quarks produced with moderate boost, $p_T\gsim 200\gev$, and remains viable for
values of the top transverse momentum up to the TeV range.
Using the results in \cite{ATLAStopres},
we find that the tagging efficiency lies in the range $5-10\%$ for
$900\lsim M_{t\bar t}\gsim 2500\gev$, where this figure includes the top-quark pair
branching fraction into $e,\mu$
semileptonic states.  This efficiency is indeed remarkedly improved over that from
traditional tagging techniques \cite{oldATLAS} (which are of order 1\%, including the
semi-leptonic branching fraction).   However, we note that
for lower values of $M_{t\bar t}$, the transitional reconstruction techniques employed here
become inefficient.  Additional sensitivity to resonance production in this range
could be gained by supplementing these techniques with traditional reconstruction methods.

We focus on the production of color-octet and color-singlet vector resonances, noting that the
color-octet case is a candidate that generates
\cite{Frampton:2009rk,Bai:2011ed} the
observed top-quark forward-backward asymmetry at the Tevatron.  The reconstruction
techniques described above
were developed for vector exchange contributions to top pair production, and some
modification would be required to account for the different acceptances associated with scalar
and tensor resonances.  Vector resonance production proceeds solely through $s$-channel
exchange in $q\bar q$ annihilation.
Contributions from gluon-gluon initial states do not occur at tree level, with
renormalizable interactions being excluded by gauge invariance.  For example,
a color octet vector resonance occurs in models with extra spatial dimensions as
a Kaluza-Klein mode of the gluon itself.  However, orthonormality of the
wave functions results in the
$g$-$g$-$g_{KK}$ coupling (where $g_{KK}$ is a gluon Kaluza Klein state)
being zero \cite{Davoudiasl:2000wi}.

The production cross section for $s$-channel vector exchange in $q\bar q\to t\bar t$ can
be written most generally as \cite{esix}
\begin{equation}
{d\sigma\over{d\cos\theta}} =  {\pi\alpha_{(s,em)}\over 2} \hat s C_F\beta \sum_{i,j}P_{ij}^{ss}
\left[ B_{ij}(1+\beta^2\cos^2\theta)+2C_{ij}\beta\cos\theta+E_{ij}(1-\beta^2)\right]\,,
\end{equation}
where the sum extends over all vector bosons being exchanged.  Here, we have defined
\begin{eqnarray}
B_{ij} & = & (v_iv_j+a_ia_j)_q(v_iv_j+a_ia_j)_t \,, \nonumber\\
C_{ij} & = & (v_ia_j+a_iv_j)_q(v_ia_j+a_iv_j)_t \,, \nonumber\\
E_{ij} & = & (v_iv_j+a_ia_j)_q(v_iv_j-a_ia_j)_t\,, \\
P_{ij}^{ss} & = & (\hat s-M_i^2)(\hat s - M_j^2)+(\Gamma_iM_i)(\Gamma_jM_j)\over
[(\hat s-M_i^2)^2+(\Gamma_iM_i)^2][(\hat s-M_j^2)^2+(\Gamma_jM_j)^2] \,, \nonumber
\end{eqnarray}
with the couplings being normalized as $g_i\bar f\gamma_\mu(v_i^f-a_i^f\gamma_5)fV_i^\mu$\,,
$\beta^2=(1-4m_t^2/\hat s)$ and $C_F$ is the usual color factor.  In what follows,
we examine the cases of a color-octet resonance with either pure vector or axial
couplings, as well as a color-singlet boson with $v^f\,, a^f$ being identical to those
of the SM $Z$ boson.  We vary the overall coupling strength and compute the width
for each case.  We assume that no new exotic decay channels contribute to the width.

\begin{figure}[htbp]
\centerline{
\includegraphics[width=9cm,angle=90]{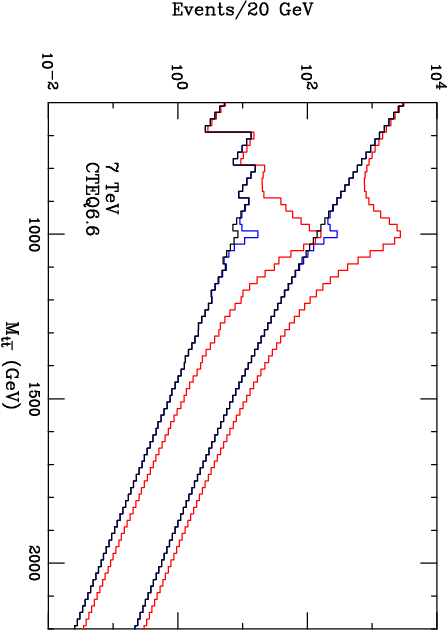}}
\vspace*{0.1cm}
\caption{Top-quark pair invariant mass spectrum including a color-octet (red) and color-singlet
(blue) vector exchange with couplings as described in the text.  The SM distribution
is repesented by the black curves.  The bottom (top) set of curves include (are without)
the semi-leptonic tagging efficiency.}
\label{bumps}
\end{figure}

Figure \ref{bumps} displays the top-quark pair invariant mass distribution for both a color-octet
and color-singlet resonance with mass of 1 TeV, as well as for the SM.  The resonance
couplings are taken to be SM strength, {\it i.e.}, $g_s$ and $g_{wk}$, respectively.
The couplings for the color-octet vector resonance are taken to be purely axial-vector,
while those for the color-singlet are taken to be the same as for the SM $Z$ boson.
The top (bottom) set of curves give the event rate without (with) including
the tagging efficiency.  The jagged curves at low invariant mass reflect the interference
between the inefficiency of the tagging routine at lower values of the top-quark $p_T$ and the
falling production cross section.  This figure demonstrates that for $M_{t\bar t}\gsim 900$ GeV, the tagging efficiencies have
indeed reached their nomimal value of $5-10\%$ stated above.\footnote{We note that the tagging efficiencies
are computed for a $M_{t\bar t}$ bin size of 100 GeV.}
We see that for the color-octet state,
the resonance is easily observable above the SM background.
However, it is clear that the resonance peak is much smaller for the color-singlet vector
boson, and hence we expect a reduced sensitivity in this case.

The recent ATLAS study \cite{ATLAStopres} demonstrated that the main source of background for
resonances in the $t\bar t$ semi-leptonic channel arise from SM top production itself, while
reducible backgrounds, such as $W+$jets, are negligible
in the $M_{t\bar t}$ range
of interest here.  Systematic errors arise from uncertainties in the NLO $t\bar t$
cross section and the parton distribution functions; we estimate that these combined
errors are of order 50\% at high
invariant masses \cite{Frederix:2007gi}.  For our discovery criteria, we require $S/\sqrt B\geq 5$ in the
$M_{t\bar t}$ region that is within 10\% of the resonance mass
on either side of the resonance peak, {\it i.e.}, $\Delta M_{t\bar t}=M_V\pm 0.1M_V$.
We include statistical as well as the 50\% theoretical systematic errors.

\begin{figure}[htbp]
\centerline{
\includegraphics[width=9cm,angle=90]{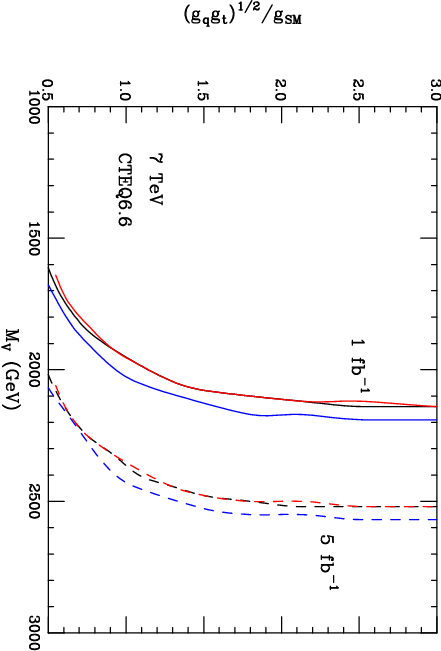}}
\vspace*{0.1cm}
\caption{Discovery reach in the coupling strength-mass plane
for purely vector (axial) coupled color-octet vector
resonances corresponding to the black (red) curves for the integrated luminosities
as indicated for the semi-leptonic channel.  The blue curves show the increase in
discovery reach when the hadronic channel is also included.}
\label{bounds}
\end{figure}

The discovery reach for a color-octet vector resonance is shown in Fig.~\ref{bounds}
in the coupling strength - mass plane for both 1 and 5 fb$^{-1}$ of integrated
luminosity at the 7 TeV LHC.  Here, we have assumed either purely vector or purely
axial vector couplings and taken $g_{SM}=g_s$.  We see that the reach is equivalent
for both types of couplings.  We see that for large couplings, the mass reach saturates
at values of $\sim 2.1(2.5)$ TeV for the two integrated luminosities.  For
smaller values of the couplings, we see that the production cross section becomes too small
to be observed once the coupling is roughly 40-50\% SM strength.  For this region,
the use of traditional top tagging techniques may enhance the discovery reach.
The corrsponding discovery reach for color-singlet vector resonances with SM-like
couplings are given in Table~\ref{zprime} for several values of the coupling strength.
Here, we see that the coupling must be roughly of order SM strength, or higher, in order
for the resonance to be observed.  These results mirror that from dijet resonance
searches \cite{Khachatryan:2010jd}.

\begin{table}
\begin{center}
\begin{tabular}{|c|c|c|} \hline\hline
$g/g_{wk}$ & 1 fb$^{-1}$ & 5 fb$^{-1}$\\ \hline \hline
1.8 & 1.50 (1.58) & 1.92 (2.00)  \\
2.0 & 1.76 (1.82) & 2.16 (2.18)\\
2.5 & 1.94 (2.02) & 2.34 (2.42)\\
3.0 & 2.12 (2.16) & 2.52 (2.56)\\
\hline\hline
\end{tabular}
\end{center}
\caption{Discovery reach in TeV for a color-singlet vector resonance with coupling
structure identical to the SM $Z$ boson for various coupling strengths relative
to $g_{wk}$ for 1 and 5 fb$^{-1}$ of integrated luminosity.  The numbers
outside (within) the parenthesis correspond to the semi-leptonic (semi-leptonic and hadronic) channel.}
\label{zprime}
\end{table}

We briefly note that in principle color-octet scalar resonances, as well as Kaluza Klein
Graviton production in models with warped extra dimensions \cite{Davoudiasl:1999jd}, can also be
observed in the $t\bar t$ invariant mass spectrum.  Most models with a heavy
scalar boson predict $q\bar qS$ couplings that are proportional to $m_q$ and hence
$s$-channel production proceeds mainly through a highly model-dependent loop level
$ggS$ 3-point function.  In addition, sizeable cross sections can be obtained in
models with warped extra dimensions via $gg\to G_{KK}\to t\bar t$ (where $G_{KK}$
is a Kaluza-Klein graviton state), however the production rates are highly
dependent on the placement of the fermions in the extra-dimensional bulk.

In order to increase event statistics, one could also include the
full hadronic channel; we provide a quick estimate of how this modifies the
discovery reach here.
In this channel we employ the HEPTopTagger of \cite{Plehn:2009rk,Plehn:2010st} discussed at length in
the previous section.  We find that the tagging efficiency lies in the range $2-4\%$ for the $M_{t\bar t}$ region of interest to us 
here, where this value includes the top-quark pair hadronic branching fraction.  An important background in this channel arises from QCD dijet production.
Requiring two subjets inside the tagged fat jets to be tagged as $b$-quarks
gives a dominant background from bottom pair production.  For the
HEPTopTagger, the probability for a QCD jet with $p_T > 200$ GeV to be
tagged as a top is approximately 2-3\%.  We roughly estimate the QCD
background using Madgraph to compute the leading order bottom pair
production cross-section, then imposing a 0.6 $b$-tagging efficiency
and a 2\% top mistag probability.  This suppresses the QCD background
beneath the dominant background of standard model top production.
The resulting effect on the discovery reach by including the hadronic channel is represented
in Fig. \ref{bounds} by the blue curves and in Table \ref{zprime} by the numbers given in parenthesis.  We see that adding this channel increases
the reach by roughly 100 GeV, indicating that this resonance search is becoming
limited by the kinematics of the parton distribution functions.


\section{Contact Operators}
\label{sec:contact}

The top-quark pair invariant mass distribution also provides constraints on models
which can generate a large forward-backward asymmetry in the absence of a narrow
resonance.  In particular, broad resonances with large couplings can easily
generate \cite{Bai:2011ed} the magnitude of the effect observed at the Tevatron.
Such non-resonant contributions to top pair production can yield significant
deviations in the invariant mass spectrum at large values
of $M_{t\bar t}$.

Numerous forms of contact operators can affect $\ttbar$ pair production at the LHC
\cite{AguilarSaavedra:2010zi,Zhang:2010dr,Degrande:2010kt,Lillie:2007hd}.
Here, we focus on a limited subset that can potentially contribute to an angular
$\ttbar$ asymmetry.
In particular, such operators must be able to interfere with the $q\bar q $-initiated
scattering process and thus we turn our attention to the usual four-quark operators.  Requiring
further that the operators  interfere with the
dominant QCD production amplitude, without chiral suppression from the light quark masses,
identifies the following set of
operators of interest:
\begin{equation}
\mc{O}_{(L/R)(L/R)}={g_{eff}^2\eta\over 2\Lambda^2}\big(
\bar q^ a\gamma ^\mu q ^ b\Big)_{L/R}\Big( \bar t^ b\gamma_\mu t ^ a\Big)_{L/R}\,,
\label{eq:4q}
\end{equation}
where the light quark flavor indices have been surpressed, $a,b$ represent color
indices, $\eta=\pm 1$,
and the subscript $L/R$
indicates that the current structure within each parenthesis can be either left- or
right-handed.  Various possible choices of the chiralities lead to different predictions
for the angular distributions for $t\bar t$ production. It is conventional to define
$g^2_{eff}=4\pi$ since the binding force is strong when
$Q^2$ approaches the cut-off $\Lambda^2$.

At the same order in $1 / \Lambda$, $\ttbar$ production also receives contributions from
two additional operators which modify the $t$-$\bar{t}$-$g$ interaction.
The chromo-magnetic operator $m_t  G^a_{\mu \nu} \bar t \sigma^{\mu\nu} \lambda^a t$
also interferes with the SM $q\bar q \to\ttbar$ production amplitude, and can affect the
$\ttbar$ production rate \cite{Atwood:1994vm}, but does not contribute to the asymmetry and
thus we do not consider it further.  The operator
$ G^a_{\mu \nu} \bar{t} \gamma^\mu \lambda^a D^\nu P_R t$ can be rewritten
using the equations of motion in terms of a vector-like, flavor-universal
four-quark operator included in Eq.~(\ref{eq:4q}) \cite{Zhang:2010dr,Lillie:2007hd}.

Interference between the contact terms and the usual gauge interactions can lead to observable
deviations from SM predictions at energies lower than the scale $\Lambda$.
Summing over the light-quark flavors, we find the following spin and color averaged
matrix element for the operators $\mc{O}_{LL,RR}$,
\beq
|\bar\mc{M} | ^ 2  = \frac{ 4\pi\eta g_s ^ 2}{ 9 \Lambda^ 2 s}\left( 
{4 m ^ 2\over s}  +
{{4 (u-m ^ 2) ^ 2}\over s^2}\right)
\eeq
and the corresponding contribution to the interference term with the SM  is
\beq
\frac{d \sigma}{d\cos\theta} = \frac{\alpha_s \beta}{9 s}\frac{\pi\eta}{2\Lambda^ 2}
           \left( {4m ^ 2\over s}+1+\beta\cos\theta+\beta^ 2\cos ^ 2\theta\right).
\eeq
For $\mc{O}_{LR, RL}$, the averaged matrix element is
\beq
|\bar\mc{M} | ^ 2  = \frac{ 4\pi\eta g_s ^ 2}{ 9 \Lambda^ 2 s}\left(
{4 m ^ 2\over s}  +
{{4 (t-m ^ 2) ^ 2}\over s^2}\right)
\eeq
giving
\beq
\frac{d \sigma}{d\cos\theta} = \frac{\alpha_s \beta}{9 s}\frac{\pi\eta}{2\Lambda^ 2}
           \left( {4m^2\over s}+1-\beta\cos\theta+\beta^ 2\cos ^ 2\theta\right).
\eeq
Note that these two sets of operators only yield different results for $t\bar t$ angular
distributions, and that after integrating over $\cos\theta$, all four operators
contribute equally to the total production rate at this order in $\Lambda$.

Summing over 20 GeV bins
in the $M_{t\bar t}$ spectrum, we employ a $\chi^2$ distribution, evaluated as usual,  to determine the 95\% C.L.
search reach for these operators.  We use only the semi-leptonic channel with
$\ell=e,\mu$ and take the tagging efficiencies as described in the previous
Section \ref{sec:resonance}.  Our results are presented in Table \ref{contactints}
for $\eta=\pm 1$ with 1 and 5 fb$^{-1}$ of the integrated luminosity at 7 TeV.   We see
that the 7 TeV LHC has a reasonable reach in the $t\bar t$ channel and that there is not
much difference between $\eta=\pm 1$.
Observation of a signal, or not, will provide information on the discovery potential
for resonance production at the higher energy 14 TeV LHC.

\begin{table}
\begin{center}
\begin{tabular}{|c|c|c|} \hline\hline
$\eta$ & 1 fb$^{-1}$ & 5 fb$^{-1}$\\ \hline \hline
 $+1$ &  3.2 (3.78) & 4.8 (7.2) \\
 $-1$ & 3.2 (3.6) &  4.7 (6.0)\\
\hline\hline
\end{tabular}
\end{center}
\caption{95\% C.L. search reach in TeV for the contact interaction scale $\Lambda$
with $\eta=\pm 1$ and 1 or 5 fb$^{-1}$ of integrated luminosity at the 7 TeV LHC.
The numbers outside (within) the parenthesis correspond to the semi-leptonic
(semi-leptonic and hadronic) event sample.} 
\label{contactints}
\end{table}

\section{Conclusions}
\label{sec:conclusion}

The LHC provides an extraordinary opportunity to study the top quark at
high energies.   Even modest amounts of LHC data afford a vision of the top quark
which is radically different from the one provided by the Tevatron, and there is much room
for discoveries and surprises.  In fact, the Tevatron may already be providing the first hints for
what the LHC may discover through its measurement of $A_{FB}^t$.  In this work, we have
examined the prospects for several early measurements of top physics at the LHC, including
remeasuring $A_{FB}^t$ itself, and searching for anomalies in the invariant mass
distribution of $t \bar{t}$ events, including resonances and contact interactions.

We have made use of the recent developments in boosted top tagging to facilitate all of these
searches.  Top-tagging is a powerful new tool that is ideal for measurements involving
energetic top quarks -- exactly the regime where early LHC data is truly probing new territory
compared to the Tevatron.  Since the measurement of $A_{FB}^t$ depends crucially
on assigning a direction to the top quark, boosted top reconstruction is a particularly incisive
tool.  Our finding is that a reasonably precise measurement of $A_{FB}^t$
is feasible with a modest amount of data, but it is worth emphasizing that
we have not fully optimized our analysis, and (particularly in light of our positive
results) more sophisticated investigation involving a full detector simulation are
warranted.
For example, the $p_T$ cut required for
tops in the measurement of $\cal{A}_F$ has not been optimized on fully showered
events, as for tops with $p_T <200\gev$ entirely different reconstruction strategies
are required.  Our results leave open the possibility that precise $\ttbar$
reconstruction combined with background rejection
at low $p_T$ could enhance prospects for measuring the asymmetry.

A recent measurement by CMS \cite{CMSAFB} takes a different approach
to the top forward-backward
asymmetry by measuring the asymmetry in the rapidity distributions of
tops and anti-tops,
$A_{C}^t = (|\eta_t |-|\eta_{\bar t} |)/(|\eta_t |+|\eta_{\bar t} |)$.
As defined, this
observable receives large contributions from the dominant $gg\to\ttbar$ process,
yielding small values for the net asymmetry.  The LHC sensitivity to
predicted values of $A_{C}^t$ are limited by
proportionally large systematic errors; if these errors can be reduced
to the percent level,
a twofold improvement over the current reported errors, then 2-3
$\sigma$ sensitivity to large BSM partonic asymmetries
may be possible in the TeV run, broadly comparable to our results.  As
our approach is statistics-limited in
the early LHC run, the comparison between our results and those of CMS
naturally raises the
question of whether it is possible to improve sensitivity at 7 TeV
with alternate reconstruction techniques.

In conclusion, we find that ${\cal A}_F (y)$ can provide an effective measurement
of the forward-backward asymmetry of top pair production at the LHC, both
in the Standard Model, as well as in models designed to explain the recent Tevatron
measurements.  Boosted top reconstruction proves a helpful tool.  We further study the
early prospects for discovery of resonances or contact interactions in $t \bar{t}$ production,
and find that with small amounts of data, the LHC will extend our knowledge to a large
degree.
The LHC is a top quark factory, and in the next years will provide a vista of top completely
unseen before now.

\acknowledgments

We gratefully acknowledge useful conversations
M.~Son, T.~Plehn, T. Rizzo, and B.~Tweedie.
T.~Tait is grateful to the SLAC theory group for their
generosity during his many visits.
The work of JH is supported by SLAC, operated
by Stanford University for the US Department of Energy under contract
DE-AC02-76SF00515.
JS was supported in part by by DOE grant DE-FG02-92ER40704, MS was supported by US
Department of Energy under contract number DE-FG02-96ER40969.  TMPT was
supported by the NSF under grant PHY-0970171.
This work was inspired in part by the workshop
``The Terascale at LHC 0.5 and Tevatron"  which was co-sponsored
by the Universities of Washington and Oregon and supported by the DOE under
contracts DE-FGO2-96-ER40956 and  DE-FG02-96ER40969 .


\end{document}